\documentclass[preprint,12pt]{elsarticle}

\usepackage{amssymb}

\journal{Journal of Computational Physics}

\begin{document}

\begin{frontmatter}



\title{
Minimizing Dispersive Errors in Smoothed Particle Magnetohydrodynamics 
for Strongly Magnetized Medium
}


\author{Kazunari Iwasaki}
\ead{kiwasaki@mail.doshisha.ac.jp}
\address{
Department of Environmental Systems Science, Doshisha University, 
Tatara Miyakodani 1-3, Kyotanabe City, Kyoto 610-0394, Japan
}

\begin{abstract}
In this study, we investigate the 
dispersive properties of smoothed particle magnetohydrodynamics (SPM)
in a strongly magnetized medium
by using linear analysis. In modern SPM, a correction 
term proportional to the divergence of the magnetic fields 
is subtracted from the equation of motion to avoid a numerical instability 
arising in a strongly magnetized medium. 
From the linear analysis, it is found that 
SPM with the correction term 
suffer from significant dispersive errors, especially for slow waves propagating 
along magnetic fields.  
The phase velocity for all wave numbers 
is significantly larger than the exact solution and has a peak at a finite wavenumber.
These excessively large dispersive errors occur because magnetic fields 
contribute an unphysical repulsive force along magnetic fields.
The dispersive errors cannot be reduced, even with
a larger smoothing length and smoother kernel functions
such as the Gaussian or quintic spline kernels.
We perform the linear analysis for this problem and find that 
the dispersive errors can be removed completely while 
keeping SPM stable if the correction term is reduced by half.
These findings are confirmed by several simple numerical experiments.
\end{abstract}


\end{frontmatter}







\section{Introduction}\label{sec:intro}
Smoothed particle hydrodynamics (SPH) is an entirely Lagrangian particle method 
for simulating fluid flows \citep{L77,GM77}.
This Lagrangian nature offers major advantages when SPH is applied to 
problems with a large, dynamic range of spatial scales. 
Furthermore, SPH can easily incorporate other physics such as self-gravity, 
radiative transfer, or chemistry. 
Thus, SPH is widely used in a variety of astrophysical problems such as 
formation of large-scale structures, galaxies, stars, and planets.

Recently, several authors have tried to extend SPH to magnetohydrodynamics (MHD) because 
magnetic fields play an important role in a variety of astrophysical environments.
In this study, we call SPH for MHD ``smoothed particle magnetohydrodynamics'' (SPM).
\citet{PM05} has developed an SPM method with artificial viscosity and resistivity 
\citep[also see][]{DS09}. 
\citet{II11} have applied Godunov's method to SPM. We call it ``GSPM''.
Recently, \citet{II15} have modified their original GSPM formulation, based on 
their derivation of the equation of motion in GSPM from an action principle.

Unfortunately, conservative formulations of SPM are known to inevitably suffer from numerical 
instability for low $\beta$ plasma because of negative stress, where $\beta$ is the 
ratio of the gas pressure to the magnetic pressure.
This instability has already been pointed out by 
\citet{PM85} \citep[also see][]{M96}. 
Among several methods proposed for removing the numerical instability \citep{PM85,M96,PM05}, 
an approach by \citet{Betal01} 
is widely used in modern SPM methods \citep[e.g.,][]{DS09,II11, II15,TP12}.
A broad discussion of stable SPM schemes is found in a review by \citet{P12}.
In the approach of \citet{Betal01}, a correction term, ${\bf B}({\bf \nabla}\cdot{\bf B})/4\pi$,
is subtracted from the right-hand side of the equation of motion.
The correction term is essentially zero if ${\bf \nabla}\cdot {\bf B}=0$ is satisfied.
By a linear analysis of SPM,
\citet{BOT04} (hereafter BOT04) found that half of the correction term,
or ${\bf B}({ \bf \nabla}\cdot{ \bf B})/8\pi$, is large enough to remove the numerical instability. 
This was confirmed by \citet{BKW12}, who found that half the 
correction term provided the least error and
minimized the violation of energy and momentum conservation in a variety of test calculations.
\citet{Betal06} have proposed a sophisticated method, wherein the size of the correction term 
varies among the SPH particles.

However, it is still unclear how the correction term affects the capability of SPM to
accurately model fluid flows.
Guiding the optimal selection of the amount of correction in a rigorous manner is important.
\citet{B95} have investigated the linear stability of various SPH formulations, kernel 
functions, and ratios of smoothing length to interparticle distance
\citep[also see][]{M96,M96phd}.
They have suggested an optimal range of parameters from obtained dispersion relations.
Although these analyses are valid only for the linear regime and a regular particle configuration, 
they provide us with a lot of knowledge for achieving further improvements to SPM schemes.
Pioneering work for SPM has been done by BOT04, who investigate the dispersive properties of SPM.
They parameterize the size of the correction term with a free parameter $\xi$ $(0\le \xi\le 1)$, and
use $\xi \times { \bf B}{ \bf \nabla}\cdot{ \bf B}/4\pi$ as the correction term. 
They found that as mentioned above, $\xi=1/2$ is large enough to stabilize SPM for low $\beta$ plasma, 
and also suggested that smoother kernels, such as Gaussian or the quintic spline kernels, reproduce 
correct phase velocities, while the cubic spline causes large dispersion errors.
However, their study is restricted to 
several linear waves in the long-wavelength limit and to the case with $\xi=1/2$
although many authors still adopt $\xi=1$.

In this study, we investigate detailed dispersive properties of SPM for low $\beta$ plasma
by changing $\xi$, the kernel functions, and the ratio of smoothing length to interparticle distance.
From the results, a suggestion of
an optimal choice of the size of the correction term is provided.

The paper is organized as follows:
in Section \ref{sec:review}, the basic equations of SPM are reviewed.
In Section \ref{sec:linear}, a dispersion relation is derived from the basic equations of SPM and 
its asymptotic behavior in the long- and short-wavelength limits is discussed.
The results of the linear analysis are presented in Section \ref{sec:result}.
To confirm the results of the linear analysis, several numerical experiments are demonstrated in 
Section \ref{sec:numerical}. 
Our results are discussed in Section \ref{sec:discuss}.
Section \ref{sec:summary} offers a summary.

\section{SPM Equations}\label{sec:review}
The basic equations of MHD are given by 
\begin{equation}
        \frac{\partial \rho}{\partial t} + { \bf \nabla}\cdot\left( \rho { \bf v} \right)=0,
        \label{eoc}
\end{equation}
\begin{equation}
        \frac{\partial v^\mu}{\partial t} = \frac{1}{\rho}\nabla^\nu T^{\mu\nu} - 
        \frac{\xi}{4\pi\rho} B^\mu \nabla^\nu B^\nu,
        \label{eom}
\end{equation}
and
\begin{equation}
        \frac{d}{dt}\left( \frac{B^\mu}{\rho} \right)
        = \frac{B^\nu}{\rho}\nabla^\nu v^\mu,
        \label{induc}
\end{equation}
where $T^{\mu\nu}$ is the stress tensor,
\begin{equation}
        T^{\mu\nu} = -\left( P+\frac{{ \bf B}^2}{8\pi} \right)\delta^{\mu\nu} 
        + \frac{B^\mu B^\nu}{4\pi},
\end{equation}
and $d/dt=\partial/\partial t + { \bf v}\cdot{ \bf \nabla}$ is the Lagrangian time derivative.
The second term on the right-hand side of equation (\ref{eom}) 
is the correction term introduced to remove the numerical instability 
(BOT04).  
The parameter $\xi$ specifies the size of the correction term.
In this study, $\xi$ is assumed to be constant for all particles.
For simplicity, instead of the energy equation, the isothermal equation of state is assumed:
\begin{equation}
        P = c^2 \rho,
\end{equation}
where $c$ is the sound speed.
In the adiabatic case, the dispersive properties of SPM are 
expected to be qualitatively the same as those in the isothermal 
case. 


In SPH, the density of the $a$-th particle is evaluated by the following equation:
\begin{equation}
        \rho_a = \sum_b m_b W({ \bf x}_a-{ \bf x}_b,h),
        \label{deni}
\end{equation}
where the subscripts denote the particle label, $m_b$ is the mass of the $b$-th particle, 
$W({ \bf x},h)$ is a kernel function, and $h$ is the smoothing length.
In the linear analysis presented in this study, the smoothing length is assumed to be 
constant. In the numerical experiments shown in Section \ref{sec:numerical}, a variable 
smoothing length is used.

There are several conservative formulations of SPM. 
Here, we show two schemes: the standard SPM formulation, 
the GSPM formulation.
The basic equations of standard SPM \citep{Betal01,PM05,DS09} are given by 
\begin{eqnarray}
    \frac{dv_a^\mu}{dt} &=& \sum_b m_b 
        \left( \frac{T_a^{\mu\nu}}{\Omega_a\rho_a^2} \nabla_a^\nu W_{ab}(h_a) + \frac{T_b^{\mu\nu}}{\Omega_b\rho_b^2}  \nabla_a^\nu W_{ab}(h_b)\right) \nonumber \\
        &-& \xi \frac{B_a^\mu}{4\pi} \sum_b m_b \left( \frac{B_a^\nu}{\Omega_a\rho_a^2} \nabla_a^\nu W_{ab}(h_a) + \frac{B_b^\nu}{\Omega_b\rho_b^2}\nabla_a^\nu  W_{ab}(h_a) \right)+ \left( \frac{d v^\mu}{dt} \right)_\mathrm{diss},
        \label{pm05 eom}
\end{eqnarray}
and
\begin{equation}
        \frac{d}{dt}\left( \frac{B^\mu}{\rho} \right)_a = 
        \frac{B_a^\nu}{\Omega_b\rho_a^2} \sum_b m_b \left( v_a^\mu - v_b^\mu \right) \nabla_a^\nu W_{ab}(h_a)
    + \left[ \frac{d}{dt} \left( \frac{B^\mu}{\rho} \right)_a\right]_\mathrm{diss},
        \label{pm05 induc}
\end{equation}
where $W_{ab}(h_a) = W({\bf x}_a - {\bf x}_b,h_a)$,
the artificial dissipation terms (viscosity and resistivity) are denoted by the subscript of ``diss'',
and $\Omega$ denotes the effect of the variation of the smoothing length.
The detailed expression is described in \citet{PM05}.

\citet{II11,II15} implemented GSPM given by 
\begin{eqnarray}
    \frac{dv_a^\mu}{dt} &=& \sum_b m_b 
        \left( \frac{ \left(T_a^{\mu\nu}\right)^*}{\Omega_a\rho_a^2} \nabla_a^\nu W_{ab}(h_a) + 
        \frac{\left(T_b^{\mu\nu}\right)^*}{\Omega_b\rho_b^2}  \nabla_a^\nu W_{ab}(h_b)\right) \nonumber \\
        &-& \xi \frac{B_a^\mu}{4\pi} \sum_b m_b \left( \frac{B_a^\nu}{\Omega_a\rho_a^2} \nabla_a^\nu W_{ab}(h_a) + \frac{B_b^\nu}{\Omega_b\rho_b^2}\nabla_a^\nu  W_{ab}(h_a) \right),
        \label{gspm eom}
\end{eqnarray}
in which the
quantities with the asterisks are the results of a Riemann solver.
The evolution equation of $B^\mu/\rho$ is the same as that of standard SPM, except 
for the artificial resistivity term. 
In \citet{II15}, the resistivity term is derived from the method of characteristics 
for Alfv{\'e}n waves \citep{SN92}.
Note that equations (\ref{gspm eom}) is a simplified version of a rigorous expression 
which is found in \citet{II11,II15}.

\citet{B95} and \citet{M96,M96phd} have shown 
that the choice of kernel functions significantly influences on dispersive properties of SPH,
and should be determined by the requirements of accuracy and computational 
efficiency.
We consider several kernel functions to
investigate how they affect the dispersive properties of SPM.
The first is the Gaussian kernel, which has the best interpolation accuracy.
It is given by 
\begin{equation}
        W_\mathrm{G}(r,h) = \left(\frac{1}{\sqrt{\pi} h}\right)^d
        e^{-(r/h)^2},
\end{equation}
where $d$ is the number of dimensions.
One disadvantage of the Gaussian kernel is its infinite extent;
in actual calculations, the contributions for $r>3.1h$ are ignored.
Also, in the linear analysis, we use a truncated Gaussian kernel at $r=3.1h$ that is denoted by $W_\mathrm{tG}$. 
Most works use the cubic spline kernel \citep{S46,ML85}, which is given by 
\begin{equation}
        W_\mathrm{4}(r,h) = \frac{C_d}{h^d} 
        \left\{
        \begin{array}{ll}
                1 - \frac{3}{2} s^2 + \frac{3}{4}s^3 & \;\;\mathrm{for}\;\; 0\le s \le 1\\
                \left( 2-s \right)^3/4 & \;\;\mathrm{for}\;\; 1< s \le 2\\
                0  & \;\;\mathrm{otherwise}
        \end{array}
        \right.,
        \label{}
\end{equation}
where $s=r/h$, and $C_1=2/3$, $C_2=10/7\pi$, and $C_3=1/\pi$.
This kernel has a great computational advantage, as it has a compact support at $r=2h$. 
However, the interpolation is inaccurate and leads to relatively large dispersion errors in
sound waves \citep{B95}.
We also consider the quintic spline kernel,
\begin{equation}
        W_\mathrm{6}(r,h) = \frac{C_d}{h^d} 
        \left\{
        \begin{array}{ll}
                -10s^5 + 30s^4 - 60s^2 + 66 & \;\;\mathrm{for}\;\; 0\le s \le 1\\
                5s^5 - 45s^4 + 150s^3 -120s^2 + 75s + 51 & \;\;\mathrm{for}\;\; 1< s \le 2\\
                -s^5 + 15s^4 -90s^3 + 270s^2 - 405 s + 243 & \;\;\mathrm{for}\;\; 2< s \le 3\\
                0  & \;\;\mathrm{otherwise}
        \end{array}
        \right.,
        \label{}
\end{equation}
where $C_1=1/120$, $C_2=7/478\pi$, and $C_3=1/120\pi$.
This kernel is smoother and more accurate kernel than $W_4$.
\section{Linear Analysis}\label{sec:linear}
A two-dimensional (2D) rectangular 
lattice of particles with an interval of $\Delta x$ is considered as an unperturbed state.
The position of the $a$-th particle position is given by ${ \bf x}_{a0} = (\Delta x) { \bf a}$, where 
${ \bf a}$ is the 2D integer vector, ${ \bf a}=(a_x,a_y)$ ($a_x,a_y=0,1,2,\cdots$).
The particle mass $m_0$ and smoothing length $h$ are assumed to be constant.
The following perturbations are considered:
\begin{equation}
        { \bf x}_a = { \bf x}_{a0} + \delta { \bf x}_a
        \label{xa}
\end{equation}
\begin{equation}
        \rho_a = \rho_0 + \delta \rho_a
\end{equation}
\begin{equation}
        { \bf v}_a = \delta { \bf v}_a
\end{equation}
\begin{equation}
        { \bf B}_a = { \bf B}_0 + \delta { \bf B}_a
\end{equation}
\begin{equation}
        P_a = P_0 + c^2\delta \rho_a,
        \label{Pa}
\end{equation}
where the subscript ``0'' indicates physical variables in the unperturbed state. 
For simplicity, the unperturbed magnetic field is assumed to be parallel to the $x$-direction, and 
fluctuations in the $z$-direction that correspond to Alfv{\'e}n wave are not considered,
although their propagation can be partly investigated by fast waves propagating 
along magnetic fields for low $\beta$ plasma.
This means that we consider fast and slow modes that oscillate in the $(x,y)$-plane.
We assume that the perturbations have the following space and time dependence:
\begin{equation}
        \delta Q_a \propto \exp\left\{ i\left( { \bf k}\cdot { \bf x}_{a0} - \omega t \right) \right\},
\end{equation}
where $\delta Q_a=(\delta{ \bf x}_a,\delta \rho_a,\delta { \bf v}_a,\delta{ \bf B}_a)$.

Substituting equations (\ref{xa})-(\ref{Pa}) into the standard 
SPM equations (\ref{deni}), (\ref{pm05 eom}), (\ref{pm05 induc}), 
and $d{ \bf x}_a/dt = { \bf v}_a$, we obtain a dispersion relation.
The artificial dissipation term is omitted in this analysis.
We do not repeat the detailed derivation of the dispersion relation that 
was already shown by BOT04. 
The dispersion relation is given by 
\begin{equation}
        \mathrm{det}\left(\omega^2 \delta^{\mu\nu} + A^{\mu\nu}\right)= 0
        \label{disp}
\end{equation}
where ``det'' indicates determinant, 
\begin{equation}
     A^{\mu\nu} = \frac{2 T_0^{\mu\eta}}{\rho_0} \left( \Psi^{\eta\nu} - \phi^\eta \psi^\nu\right)
     - \frac{1}{\rho_0}\left\{P_0\phi^\mu\psi^\nu
     + \frac{1}{4\pi}\left( { \bf B}_0^2 \delta^{\mu\nu} - B_0^\mu B_0^\nu\right)
     {\phi}^\zeta{\psi^\zeta}\right\},
        \label{Aij}
\end{equation}
$T_0^{\mu\nu}$ is the unperturbed stress tensor modified by the correction term,
\begin{equation}
    T_0^{\mu\nu} = - \left( P_0 + \frac{{ \bf B}_0^2}{8\pi} \right)\delta^{\mu\nu} 
    + \left( 1-\xi \right)\frac{B_0^\mu B_0^\nu}{4\pi},
\end{equation}
$\psi^\mu$, $\phi^\mu$, and $\Psi^{\mu\nu}$ are given by 
\begin{equation}
        \psi^\mu = \sum_{b} \frac{m_0}{\rho_0}\left( 1- e^{-i{ \bf k}\cdot \left( { \bf x}_{a0} - { \bf x}_{b0}\right)} \right)
        \frac{\partial W_{ab,0}}{\partial x_{a0}^\mu},
        \label{psi}
\end{equation}
\begin{equation}
        \phi^\mu = \sum_{b} \frac{m_0}{\rho_0}\left( 1 + e^{-i{ \bf k}\cdot \left( { \bf x}_{a0} - { \bf x}_{b0}\right)} \right)
        \frac{\partial W_{ab,0}}{\partial x_{a0}^\mu},
        \label{phi}
\end{equation}
and
\begin{equation}
        \Psi^{\mu\nu} = \sum_{b} \frac{m_0}{\rho_0}\left( 1 - e^{-i{ \bf k}\cdot \left( { \bf x}_{a0} - { \bf x}_{b0}\right)} \right)
        \frac{\partial^2 W_{ab,0}}{\partial x_{a0}^\mu\partial x_{a0}^\nu}.
        \label{Psi}
\end{equation}
The first term on the right-hand side of equation (\ref{Aij}) comes from the perturbation of 
$(1/\rho_a^2 + 1/\rho_b^2)\nabla^\nu W_{ab}$, and the second term comes from the perturbation of $T^{\mu\nu}$ in equation (\ref{pm05 eom}).

Note that the dispersion relation derived from the linearized GSPM equations is 
identical to equation (\ref{disp}) if
the physical quantities with an asterisk are evaluated at the arithmetic mean 
between the $a$- and $b$-th particles.
Thus, this linear analysis is valid both in standard SPM and GSPM.

Since equation (\ref{disp}) is a quadratic equation in terms of $\omega^2$, 
two modes are obtained. In this study, 
the mode with larger (smaller) $\omega^2$ is referred to as the fast (slow) mode. 
The numerical phase velocities  $\omega/|{ \bf k}|$ 
of the fast and slow modes are denoted by $c_\mathrm{f,num}({ \bf k})$ and 
$c_\mathrm{s,num}({ \bf k})$, respectively.

\subsection{Asymptotic behaviors}\label{sec:asym}
In this section, we investigate the asymptotic behavior 
of the dispersion relation in the short- and long-wavelength limits.

\subsubsection{Short-wavelength Limit}\label{sec:shortlimit}
SPM without the correction term 
is unstable for low $\beta$ plasma because negative diagonal components
appear in the stress tensor \citep{M96}.
The development of numerical instability begins with the growth of fluctuations
whose scales are comparable to $\Delta x$.
Thus, in this section, we investigate the asymptotic behavior in the short-wavelength limit and 
the size of $\xi$ required to stabilize SPM.
Note that this already has been done by BOT04, who numerically solve the dispersion relation.
Here, we show that their conclusion is reproduced in the following simple analytical manner.

In the discrete system, the largest wavenumber is $k = \pi/\Delta x$ where
the unit wavelength is expressed by two particles.
A compressible wave propagating along ${ \bf B}_0$ (${ \bf k}=(\pi/\Delta x,0)$) is considered.
This corresponds to the slow mode.
In this case, we obtain $\phi^\mu=\psi^\mu=0$ from equations (\ref{psi}) and (\ref{phi}).
From equation (\ref{disp}), the dispersion relation becomes
\begin{equation}
        \omega^2 = -A^{xx} = \frac{2P_0}{\rho_0}\left( 1 - \frac{1-2\xi}{\beta} \right)
        \sum_b \frac{m_0}{\rho_0} \left( 1-\left( -1 \right)^{a_x-b_x} \right)
        \frac{\partial^2 W_{ab,0}}{\partial x_{a0}^2},
        \label{shorteq}
\end{equation}
where $\beta \equiv 8\pi P_0/B_0^2$.
The summation in equation (\ref{shorteq}) is positive in normal situations \citep{M96}.
Without the correction term ($\xi=0$), one can see that SPM becomes unstable, 
since $\omega^2<0$ if $\beta<1$. 
To be stable, the following condition should be satisfied:
\begin{equation}
        \xi > \xi_\mathrm{min} = \frac{1}{2}\left( 1-\beta \right),
        \label{criterion}
\end{equation}
where $\xi_\mathrm{min}$ is the minimum value of $\xi$ needed to ensure stability for a given $\beta$.
This linear dependence of $\xi_\mathrm{min}$ on $\beta$ is 
the same as that found numerically in BOT04 (see their Fig. 7).
In the strong magnetic field limit $(\beta\rightarrow 0)$, 
$\xi=1/2$ is large enough to stabilize SPM.

\subsubsection{Long-wavelength Limit}\label{sec:longlimit}
In the long-wavelength limit ($|{ \bf k}|\Delta x\rightarrow0$), 
summations can be replaced by integrals in equation (\ref{disp}) \citep{M89}.
For instance, $\psi^\mu$ is approximated by 
\begin{equation}
        \psi^\mu \sim \int 
        \left( 1- e^{-i{ \bf k}\cdot { \bf x}} \right)\frac{\partial W({ \bf x},h)}{\partial x^\mu}d^2 x 
        = - ik^\mu \hat{W}({ \bf k}),
        \label{psilong}
\end{equation}
where integration by parts and $W\rightarrow 0$ for $x\rightarrow\infty$ are used, and 
$\hat{W}({ \bf k})$ is the Fourier transform of $W({ \bf x},h)$, given by 
\begin{equation}
  \hat{W}({ \bf k},h) \equiv  \int e^{-i{ \bf k}\cdot { \bf x}} W({ \bf x},h) d^2 x.
\end{equation}
In the similar way, one obtains
\begin{equation}
        \phi^\mu \sim i k^\mu \hat{W}\;\;\mathrm{and}\;\;
        \Psi^{\mu\nu} \sim k^\mu k^\nu \hat{W}.
    \label{philong}
\end{equation}
Using equations (\ref{psilong}) and (\ref{philong}), equation (\ref{Aij}) becomes
\begin{equation}
        A^{\mu\nu} \sim  \frac{2}{\rho_0} T_0^{\mu\eta}k^\eta k^\nu \hat{W}\left( 1 - \hat{W} \right)
        - \frac{1}{\rho_0}\left\{ P_0k^\mu k^\nu + 
        \frac{1}{4\pi}\left( { \bf B}_0^2 \delta^{\mu\nu} - B_0^\mu B_0^\nu\right)k^2 \right\}
         \hat{W}^2.
        \label{Aijideal}
\end{equation}
If the Gaussian kernel is applied, its Fourier transform is also Gaussian, 
$\hat{W}({ \bf k},h)=e^{-h^2k^2/4}$. Thus, $\hat{W}\rightarrow 1 + O( (hk)^2)$ 
for $|{ \bf k}|h\rightarrow 0$.
Using this fact, the dispersion relation (equation (\ref{disp})) becomes
\begin{equation}
\omega^2 \sim k^2 \left\{
c^2 + \frac{{ \bf B}_0^2}{4\pi\rho_0}
\pm \sqrt{\left(  c^2 + \frac{{ \bf B}_0^2}{4\pi\rho_0}\right)^2 
- c^2\frac{({ \bf B}_0\cdot { \bf k})^2}{\pi \rho_0 k^2} 
} 
\right\}.
        \label{dispideal}
\end{equation}
This dispersion relation holds for both fast and slow waves.
Note that equation (\ref{dispideal}) does not contain $\xi$. Thus, 
waves in the long-wavelength limit are not affected by the correction term.
Ideally, the dispersion relation of SPM satisfies the correct phase velocity as long as 
equations (\ref{psilong}) and (\ref{philong}) are valid.

Comparing equations (\ref{Aijideal}) and (\ref{Aij}), one can see that 
the correct phase velocities come from the second terms on the right-hand sides. 
The first term on the right-hand side of equation (\ref{Aijideal}) becomes zero 
because $\hat{W}\sim 1$.
In reality, the numerical dispersion relation (\ref{disp}) is expected to 
deviate from that in equation (\ref{dispideal}) because finite discretization errors are introduced 
in equations (\ref{psilong}) and (\ref{philong}).
The first term on the right-hand side of equation (\ref{Aij}) 
causes larger dispersive errors than the second term.
This is because $\Psi^{\mu\nu}$ contains the second derivative of the kernel function 
that has larger errors (see equation (\ref{Psi})).
In particular, the cubic spline kernel leads to large errors in $\Psi^{\mu\nu}$ because its second 
derivative is a broken line.
Thus, dispersive errors mainly come from the first term on the 
right-hand side of equation (\ref{Aij}).

\section{Results}\label{sec:result}
\subsection{Phase Velocities in Long-wavelength Limit}\label{sec:long}
In this section, the dispersion relation (\ref{disp}) is solved by considering a sufficiently 
small wavenumber $(|{ \bf k}|=10^{-3}\pi/\Delta x)$ and 
changing the angle $\theta$ between ${ \bf k}$ and ${ \bf B}_0$ 
in order to investigate whether SPM correctly reproduces the phase velocities shown in Section \ref{sec:longlimit}.

\subsubsection{Fast Mode}\label{sec:long fast}

\begin{figure}
        \begin{center}
                \includegraphics[width=13cm]{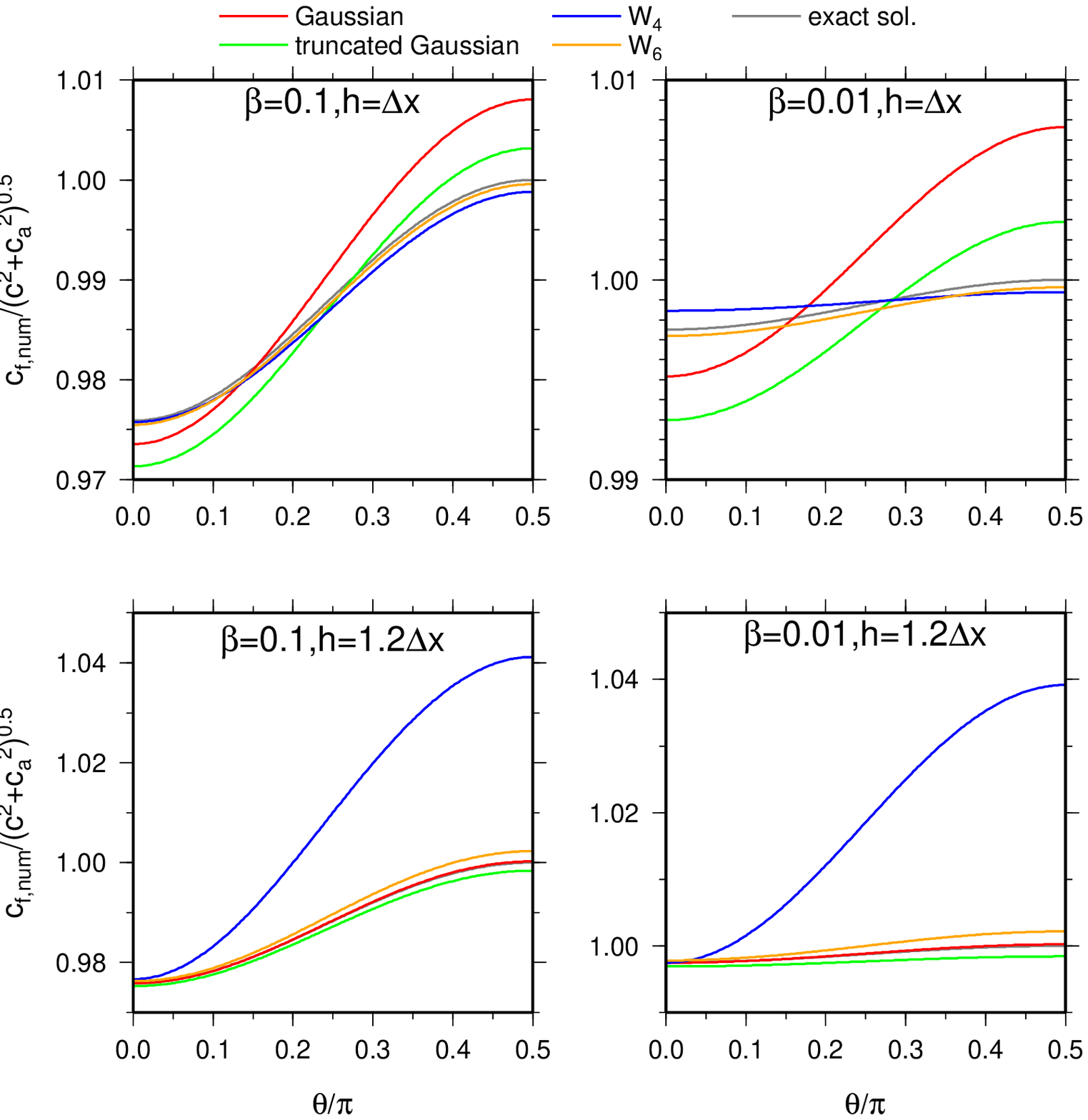}
        \end{center}
        \caption{
        Phase velocity of the fast mode versus the angle between ${ \bf k}$ and ${ \bf B}_0$ 
        for $(\beta,h)=(0.1,\Delta x)$, $(0.01,\Delta x)$, $(0.1,1.2\Delta x)$, and $(0.01,1.2\Delta x)$. 
        The results with $\xi=1$ and $\xi=1/2$ are almost identical, and 
        so only those with $\xi=1$ are presented. 
        The red, green, blue, and orange lines correspond to the results using the Gaussian, 
        truncated Gaussian,
        cubic spline, and quintic spline kernels. In each panel, 
        the exact solutions are shown by the gray line.
        }
        \label{fig:fdiag xi1.0}
\end{figure}

Fig. \ref{fig:fdiag xi1.0} shows the numerical phase velocities of the fast mode as a function of the angle 
$\theta$ for various values of $\beta$, $h$, and various kernel functions.
The value of $\xi$ is assumed to be 1.
The exact solutions are plotted by the gray lines.
For low $\beta$ plasma, the fast wave at $\theta=0$ represents the 
incompressible pseudo-Alfv{\'e}n wave whose 
phase velocity is $c_\mathrm{a}\equiv B_0/\sqrt{4\pi\rho_0}$.  
The phase velocity increases with $\theta$ and gradually changes into a compressible mode.
At $\theta=\pi/2$, the phase velocity reaches a maximum value of $\sqrt{c^2+c_\mathrm{a}^2}$.
In the upper column of Fig. \ref{fig:fdiag xi1.0}, for $h=\Delta x$, 
all kernel functions reproduce the exact solutions within an error of 1\% although
the Gaussian kernel provides slightly worse results. For $h=1.2\Delta x$, 
the results for the Gaussian kernel 
are almost identical to the exact solution (see the lower column of Fig. \ref{fig:fdiag xi1.0}).
Only the results obtained with $W_4$ suffer from relatively large errors.
The results for $\xi=1/2$ are not shown, because
it is confirmed that the numerical dispersion relations do not much depend on $\xi$. 

It is found that 
SPM can reproduce the phase velocity of the fast mode reasonably well regardless of $\xi$.
This behavior can be qualitatively 
understood from the one-dimensional dispersion relation for $k_x=0$ ($\theta = \pi/2$) given by 
\begin{equation}
\omega^2 = \left( c^2 + 2c_\mathrm{a}^2 \right)\left( \Psi^{yy} - \phi^y \psi^y \right)
+ \left( c^2 + c_\mathrm{a}^2 \right) \phi^y \psi^y. 
  \label{fast1dim}
\end{equation}
As mentioned in Section \ref{sec:longlimit}, 
the correct phase velocity $\left(\sqrt{c^2 + c_\mathrm{a}^2}\right)$ 
comes from the second term on the right-hand side of equation (\ref{fast1dim}) and 
the dispersion errors arise mainly from the first term.
To obtain the correct phase velocity, the first term on the right-hand side of 
equation (\ref{fast1dim}) should
be negligible compared with the second term. 
We can see that the coefficient of $(\Psi^{yy}-\phi^y\psi^y)$ in
equation (\ref{fast1dim}) is comparable to that 
of $\phi^y\psi^y$.
Thus, as long as $|(\Psi^{yy}- \phi^y\psi^y)/(\phi^y\psi^y)|\ll 1$ is satisfied, 
the numerical phase velocities agree with the exact values within sufficiently small errors.
For $\theta \ne \pi/2$, a term proportional to $\xi$ appears in the first term on the 
right-hand side of equation (\ref{fast1dim}).
Also in this case, the effect of the first term can be neglected, as the 
phase velocity of the fast mode is comparable to $c_\mathrm{a}$. 
That is why the numerical phase velocities  agree with the exact phase velocities reasonably well and 
do not depend much on $\xi$ for all $\theta$.

\subsubsection{Slow Mode}\label{sec:long slow}
\begin{figure}
        \begin{center}
                \includegraphics[width=13cm]{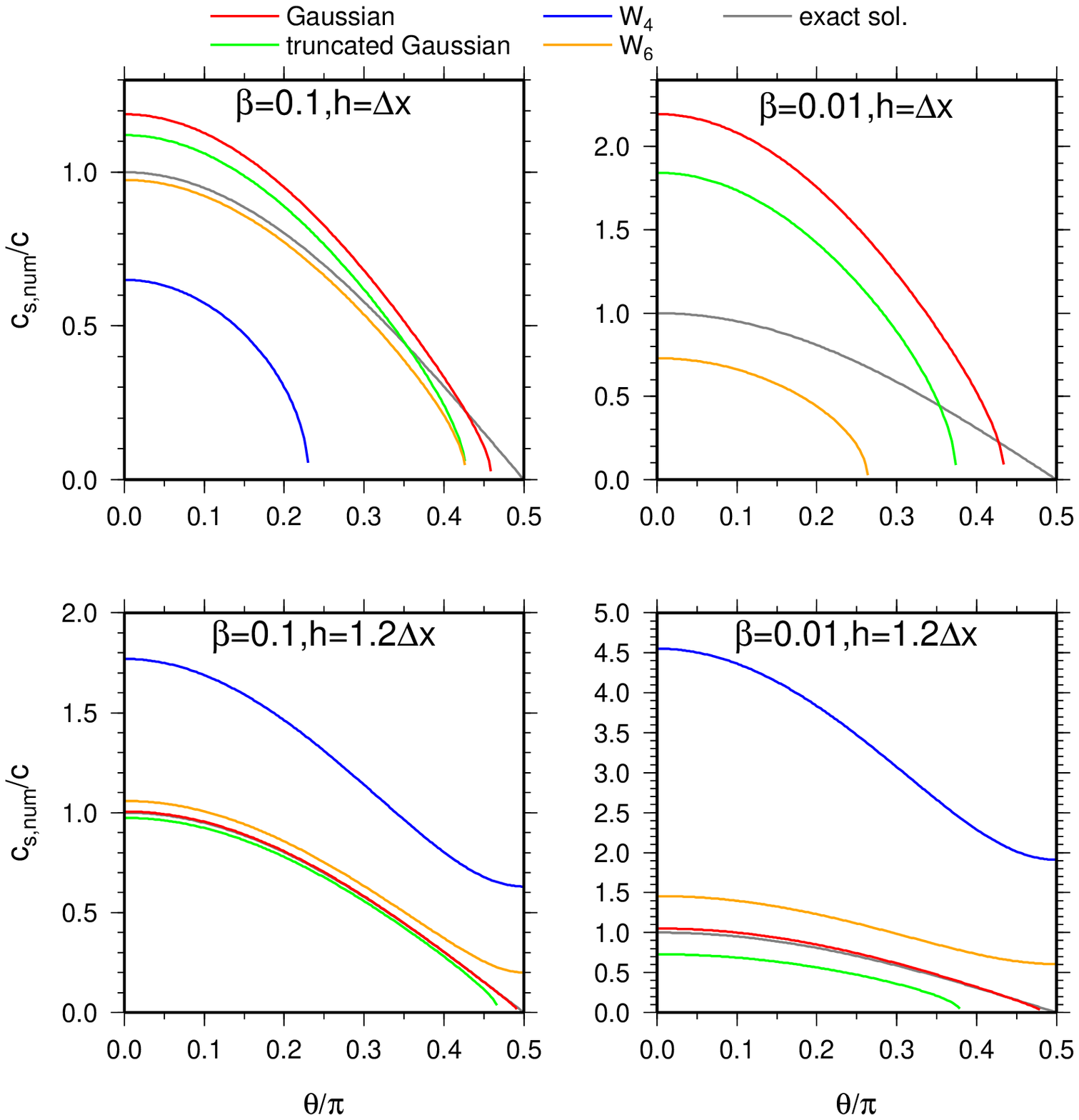}
        \end{center}
        \caption{
        Phase velocity of the slow mode versus the angle between ${ \bf k}$ and ${ \bf B}_0$ 
        for $(\beta,h)=(0.1,\Delta x)$, $(0.01,\Delta x)$, $(0.1,1.2\Delta x)$, and $(0.01,1.2\Delta x)$. 
        The red, green, blue, and orange lines correspond to the results using the Gaussian, 
        truncated Gaussian,
        cubic spline, and quintic spline kernels. In each panel, 
        the exact solutions are shown by the gray line.
        All results are taken at $\xi=1$.
        }
        \label{fig:diag xi1.0}
\end{figure}
The slow mode exhibits completely different dispersion properties than the fast mode.
Fig. \ref{fig:diag xi1.0} shows the numerical phase velocities of the slow mode 
as a function of the angle $\theta$ for various values of $\beta$ and $h$, and various
kernel functions, with $\xi=1$.
At $\theta =0$, the slow mode corresponds to the sound wave propagating in the direction parallel to ${ \bf B}_0$.
The phase velocity decreases with $\theta$ and gradually becomes the incompressible mode. At $\theta=\pi/2$, the 
phase velocity becomes zero.
First, we focus on the cases with $h=\Delta x$ shown in the upper panels of Fig. \ref{fig:diag xi1.0}.
The results with $W_4$ are significantly underestimated for $\beta=0.1$ and 
leads to a numerical instability at $\beta=0.01$.
Even for the smoother kernels ($W_\mathrm{G}$, $W_\mathrm{tG}$, $W_6$), the dispersion errors are significantly large 
and becomes worse for smaller $\beta$.
Although only $W_6$ provides a better result for $\beta=0.1$, the stronger magnetic field ($\beta=0.01$) makes the 
results worse.
From the lower left panel in Fig. \ref{fig:diag xi1.0}, for $\beta=0.1$, 
larger $h$ ($h=1.2\Delta x$) improves the results with the smoother kernels while
the result with $W_4$ does not improve.
Also, in the case with larger $h$, for the stronger magnetic field ($\beta=0.01$), 
the dispersion errors become worse except for $W_\mathrm{G}$.
In summary, the dispersion errors of the slow mode are large and becomes significant as $\beta$ decreases. 
A larger smoothing length makes the errors lower with every kernel except for the cubic spline, although
large errors still remain for sufficiently low $\beta$.
Thus, larger smoothing length cannot be an ultimate solution.


\begin{figure}
        \begin{center}
                \includegraphics[width=13cm]{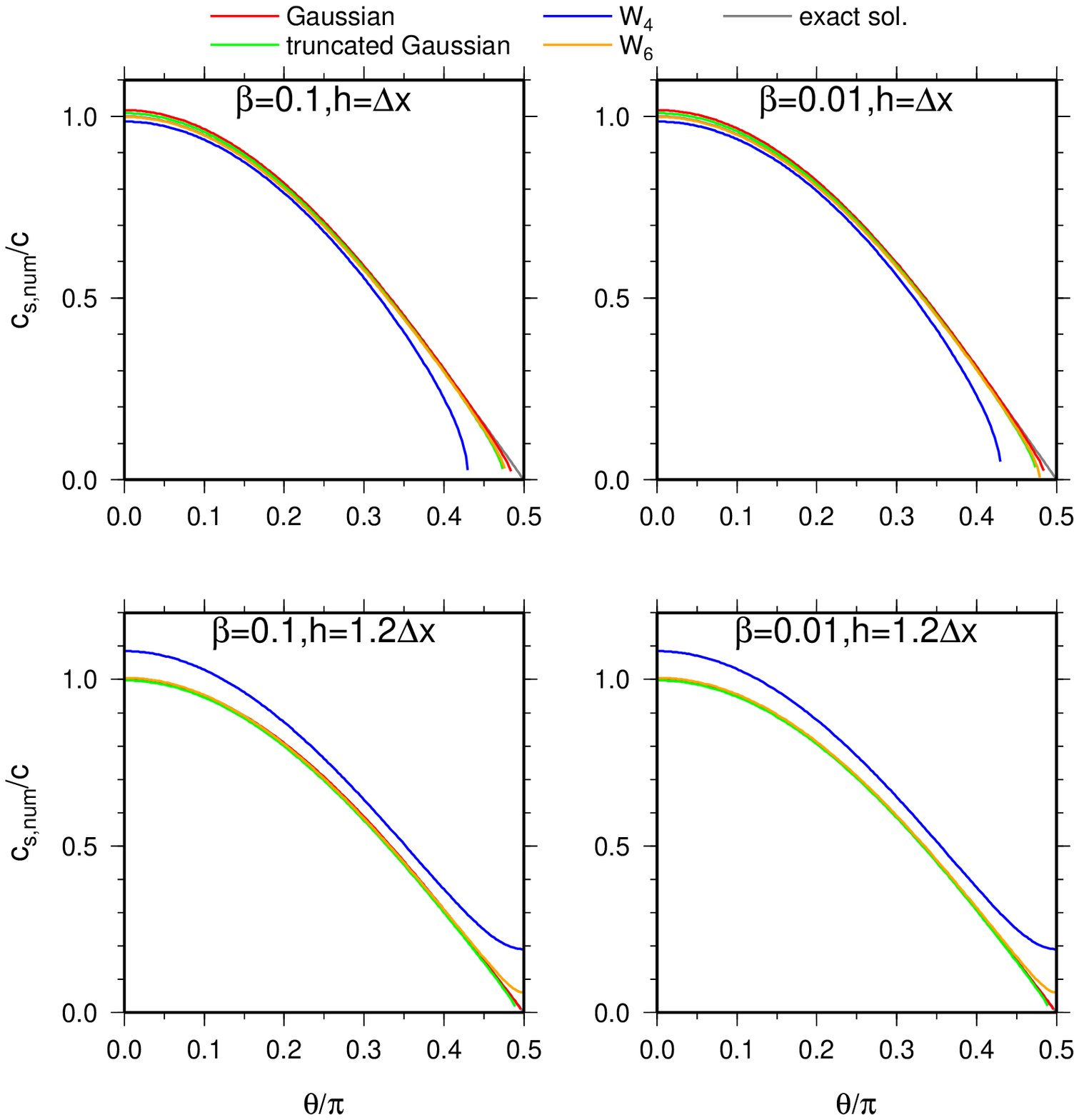}
        \end{center}
        \caption{
        Phase velocity of the slow mode versus the angle between ${ \bf k}$ and ${ \bf B}_0$ 
        for $(\beta,h)=(0.1,\Delta x)$, $(0.01,\Delta x)$, $(0.1,1.2\Delta x)$, and $(0.01,1.2\Delta x)$. 
        The red, green, blue, and orange lines correspond to the results using the Gaussian, 
        truncated Gaussian,
        cubic spline, and quintic spline kernels. In each panel, 
        the exact solutions are shown by the gray line.
        All results are taken at $\xi=1/2$.
        }
        \label{fig:diag xi0.5}
\end{figure}
Fig. \ref{fig:diag xi0.5} is the same as Fig. \ref{fig:diag xi1.0} but setting $\xi=1/2$.
All kernels can provide the correct phase velocities for all cases although 
the results with $W_4$ have relatively large errors.
BOT04 have already investigated the dispersion properties for $\xi=1/2$,
and their results are consistent with ours.

Figs. \ref{fig:diag xi1.0} and \ref{fig:diag xi0.5} reveal that SPM with $\xi=1$ exhibits
significantly large dispersive errors that disappear if $\xi=1/2$ is used.
This behavior can be qualitatively 
understood from the simple one-dimensional dispersion relation for $k_y=0$ ($\theta = 0$) 
given by 
\begin{equation}
\omega^2 = \left\{ 2c^2 + c_\mathrm{a}^2(2\xi-1) \right\}
   \left( \Psi^{xx} - \phi^x \psi^x \right)
   + c^2 \phi^x \psi^x. 
  \label{slow1dim}
\end{equation}
To obtain the correct phase velocity, $c$, the first term on the right-hand side of 
equation (\ref{slow1dim}) should be negligible when compared with the second term.
However, in contrast to the fast mode case, 
the coefficient of $(\Psi^{xx}-\phi^x\psi^x)$ in equation (\ref{slow1dim}) 
is much larger than that of $\phi^x\psi^x$ for low $\beta$ ($c_\mathrm{a}\gg c$).
Thus, the first term can be important even if $|(\Psi^{xx}-\phi^x\psi^x)/(\phi^x\psi^x)|$ is sufficiently small.
Note that the contribution of the magnetic field disappears only at $\xi=1/2$ in the first term 
of equation (\ref{slow1dim}), and the dispersion relation is reduced to that without a magnetic field.
That is why the errors are mostly eliminated for $\xi=1/2$, as shown in Fig. \ref{fig:diag xi0.5}. 

These findings in the long-wavelength limit suggest that the best choice for $\xi$ is 1/2, as
SPM with $\xi = 1$ suffers from significant errors.

\subsection{Dispersion Relation}\label{sec:disp}
In this section, the overall dispersive properties of SPM are investigated.
\begin{figure*}
        \begin{center}
                \includegraphics[width=14cm]{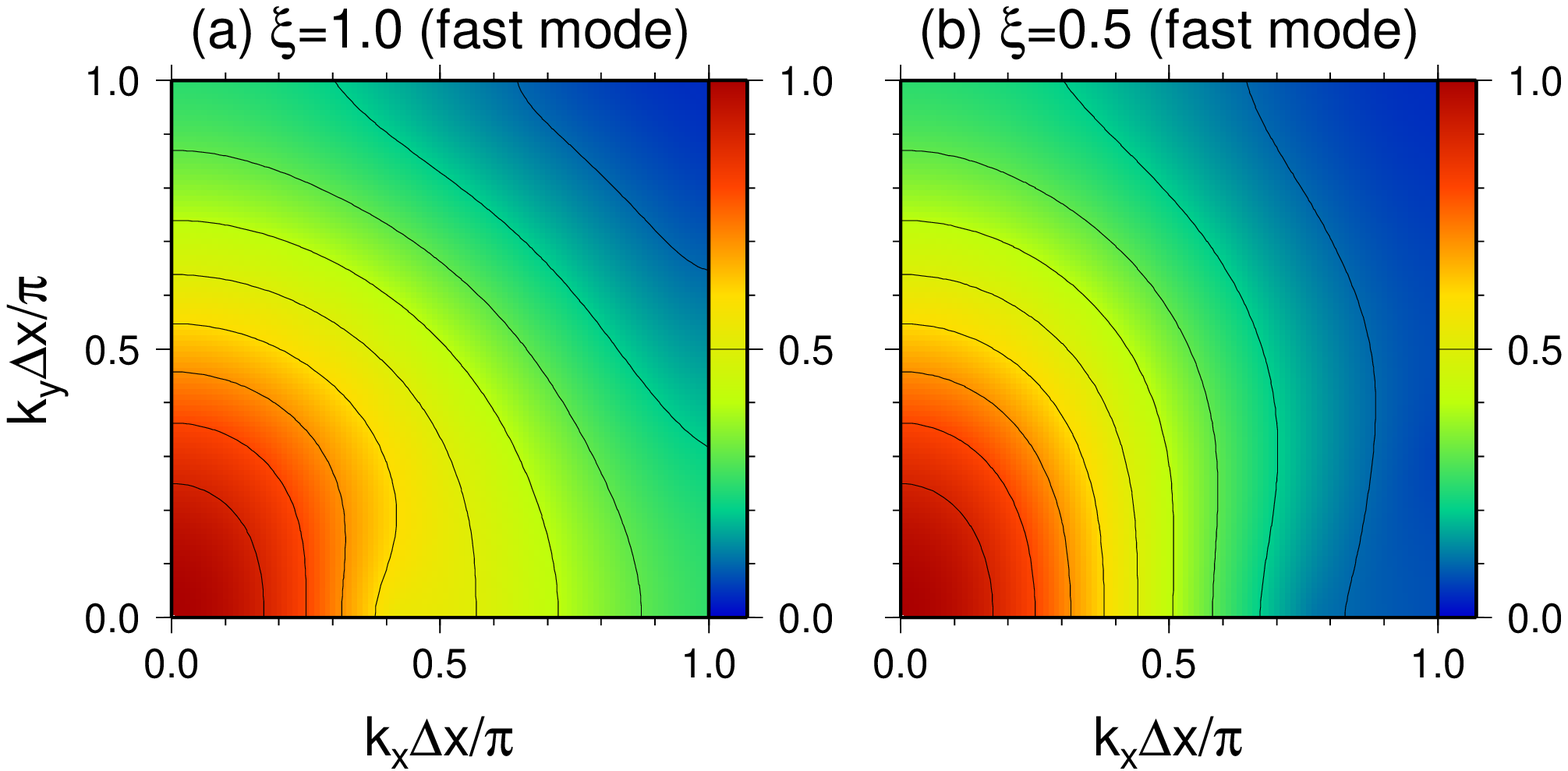}
        \end{center}
        \caption{
        Color maps of the phase velocities of the fast mode in the 
        $(k_x \Delta x/\pi,k_y \Delta x/\pi)$ plane for 
        (a)$\xi=1$ and (b)$\xi=1/2$.
        The Gaussian kernel is considered.
        The parameters $\beta=0.1$ and $h=1.2\Delta x$ are used.
        In each panel, the color shows the numerical phase velocity normalized by the exact phase velocity 
        depending on $\theta$.
        The gray region corresponds to the unstable region.
        }
        \label{fig:kxyf}
\end{figure*}
Figs. \ref{fig:kxyf}a and \ref{fig:kxyf}b show color maps of the numerical phase velocities of the fast mode in the 
$(k_x\Delta x/\pi,k_y\Delta x/\pi)$ plane for $\xi=1$ and $1/2$, respectively.
The smoothing length is assumed to be $h=1.2\Delta x$, and the plasma $\beta$ value 
is fixed to be $\beta=0.1$.
Here, we focus on the results with the Gaussian kernel. It is confirmed that 
the qualitative properties do not depend on the kernel functions.
In Fig. \ref{fig:kxyf}, the numerical phase velocities are divided by the corresponding 
exact solutions depending 
on $\theta$.
In both panels in Fig. \ref{fig:kxyf}, 
$c_\mathrm{f,num}/c_\mathrm{f}(\theta)$ has peaks at the origin and monotonically 
decreases with $|{ \bf k}|$ at all $\theta$, where $c_\mathrm{f}$ is the exact phase velocity of the fast mode.
The difference between the results with $\xi=1$ and $1/2$ is found only in the region where 
$|{ \bf k}|\Delta x/\pi>0.5$ and $\theta\sim 0$.
For $\xi=1/2$, $c_\mathrm{f,num}/c_\mathrm{f}$ declines more rapidly than it does for $\xi=1$.
This behavior will be explained later.
Thus, the dispersion relation of the fast mode does not depend much on the value of $\xi$.
This is consistent with the findings in Section \ref{sec:long fast}.

\begin{figure*}
        \begin{center}
                \includegraphics[width=14cm]{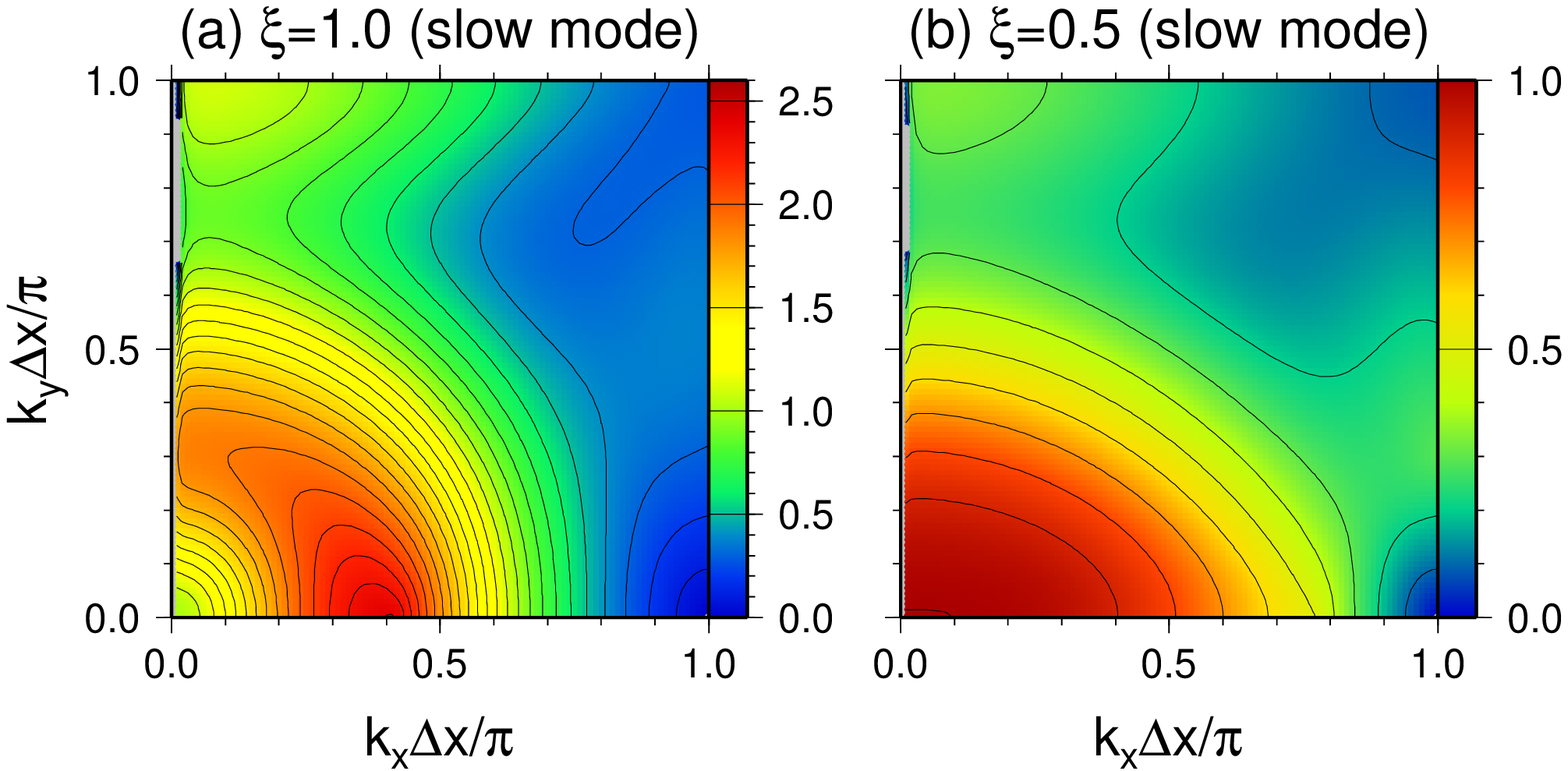}
        \end{center}
        \caption{
        Color maps of the phase velocities of the slow mode in the 
        $(k_x \Delta x/\pi,k_y \Delta x/\pi)$ plane for 
        (a)$\xi=1$ and (b)$\xi=1/2$.
        The Gaussian kernel is considered.
        The parameters $\beta=0.1$ and $h=1.2\Delta x$ are used.
        In each panel, the color shows the numerical phase velocity normalized by the exact phase velocity 
        depending on $\theta$.
        The gray region corresponds to the unstable region.
        }
        \label{fig:kxys}
\end{figure*}
Next, the dispersion relations of the slow mode are investigated.
Fig. \ref{fig:kxys} is the same as Fig. \ref{fig:kxyf} but for the slow mode.
First, the result with $\xi=1$, as shown in Fig. \ref{fig:kxys}a, is investigated.
Fig. \ref{fig:kxys}a shows that $c_\mathrm{s,num}/c$ exhibits anomalous features,
although $c_\mathrm{s,num}$ can reproduce the correct phase velocities in the 
long-wavelength limits for the Gaussian kernel (see the lower left panel in Fig. \ref{fig:diag xi1.0}). 
$c_\mathrm{s,num}/c_\mathrm{s}$ increases from the origin with $|{ \bf k}|$ especially in the direction 
parallel to ${ \bf B}_0$ ($\theta \sim 0$).
Around $|{ \bf k}| \Delta x/\pi \sim 0.4$, $c_\mathrm{s,num}/c_\mathrm{s}$ reaches a maximum whose value 
is $\sim 2.5$ times larger than the exact solution. 
For larger $|{ \bf k}|$ $(|{ \bf k}|\Delta x/\pi>0.4$), $c_\mathrm{s,num}$ decreases with 
$|{ \bf k}|$. 

These anomalous features of the dispersion relation for $\xi=1$ completely disappear 
at $\xi=1/2$, as shown in Fig. \ref{fig:kxys}b where
$c_\mathrm{s,num}/c_\mathrm{s}$ monotonically decreases with $|{ \bf k}|$ from the center.

\begin{figure*}
        \begin{center}
                \includegraphics[width=14cm]{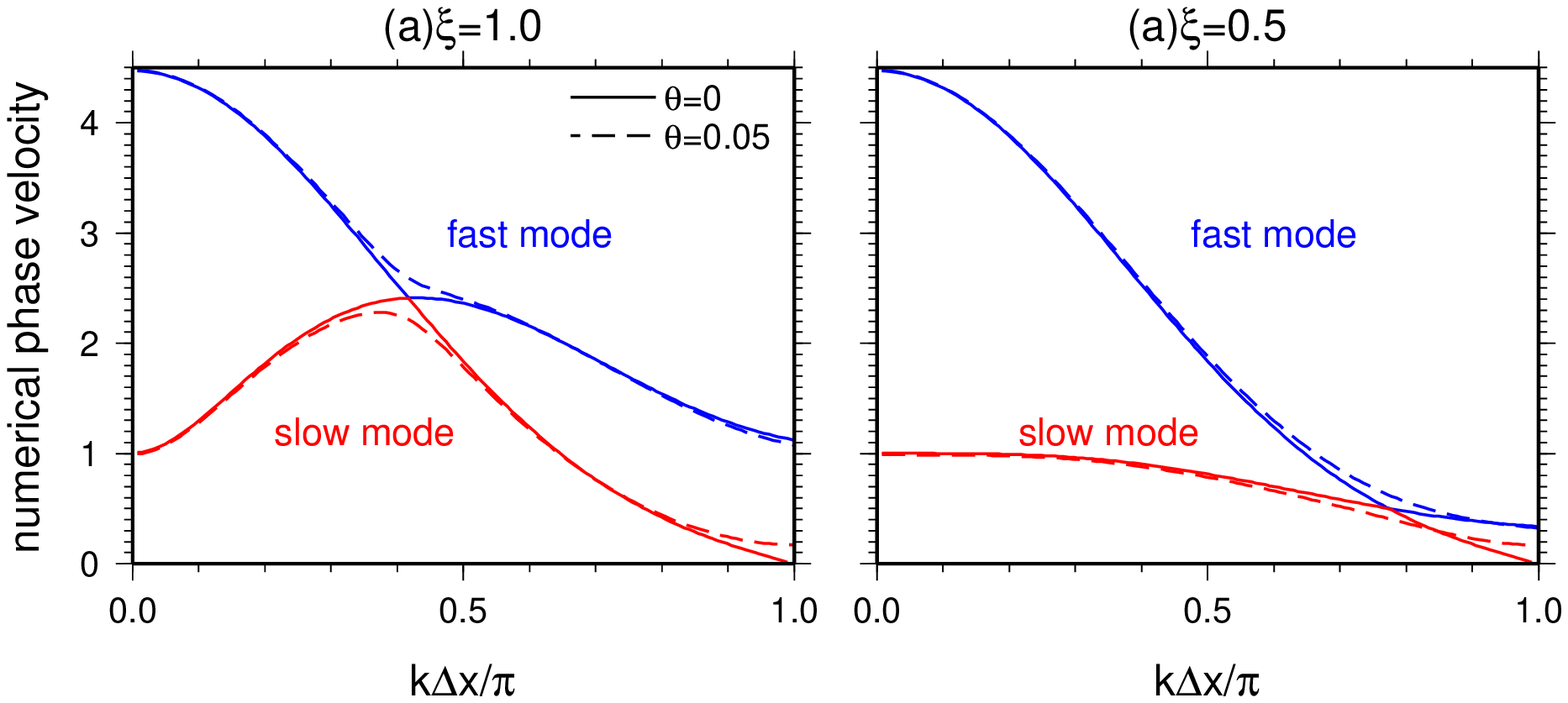}
        \end{center}
        \caption{
        Dispersion relations for (a)$\xi=1$ and (b)$\xi=1/2$.
        The Gaussian kernel is used.
        In each panel, the blue and red lines corresponds to the fast and slow modes, respectively.
        The solid and dashed lines indicate the results with $\theta=0$ and $\theta=0.05$, respectively.
        }
        \label{fig:cross}
\end{figure*}

Fig. \ref{fig:kxys} reveals that the phase velocities around $\theta=0$ strongly depend on $\xi$.
To see this more clearly, the phase velocities at $\theta = 0$ and $\theta=0.05$
are plotted in Fig. \ref{fig:cross} as a function of $|{ \bf k}|$.
Fig. \ref{fig:cross}a shows the results with $\xi=1$.
First, we focus on the case where $\theta=0$, shown by the solid lines.
As mentioned above, the numerical phase velocities agree with the exact values for both 
the fast and slow modes.
From $k=0$, $c_\mathrm{f,num}$ decreases with $k$ while $c_\mathrm{s,num}$ increases.
At $k\Delta x/\pi\sim 0.4$, the two branches have the same phase velocity. 
Beyond this point, it can be clearly seen that the fast mode branch 
smoothly connects with the slow mode branch. 
This happens simply because larger (smaller) $\omega^2$ is 
referred to as the fast (slow) mode in this study, regardless of its eigenfunction.
The ``real'' fast (slow) mode should be an incompressible (compressible) mode at $\theta=0$.
By investigating the corresponding eigenfunctions, it is confirmed that the ``real'' fast and slow modes 
intersect at $k\Delta x/\pi\sim 0.4$. This means that the slow mode changes into an 
incompressible mode and the fast mode changes into a compressible mode beyond the intersection point.
The dashed line in Fig. \ref{fig:cross} indicates the results with $\theta = 0.05$.
We can see the mode exchange between the fast and slow branches around $k\Delta x/\pi\sim 0.4$.
From the eigenfunctions, it is confirmed that the fast (slow) branch is changed into a compressible 
(incompressible) mode by the mode exchange.
Fig. \ref{fig:cross}a indicates 
that the phase velocity of the compressible mode (the ``real'' slow mode) is supersonic $(>c_\mathrm{s})$ for all wavenumbers. 

\begin{figure*}
        \begin{center}
                \includegraphics[width=7cm]{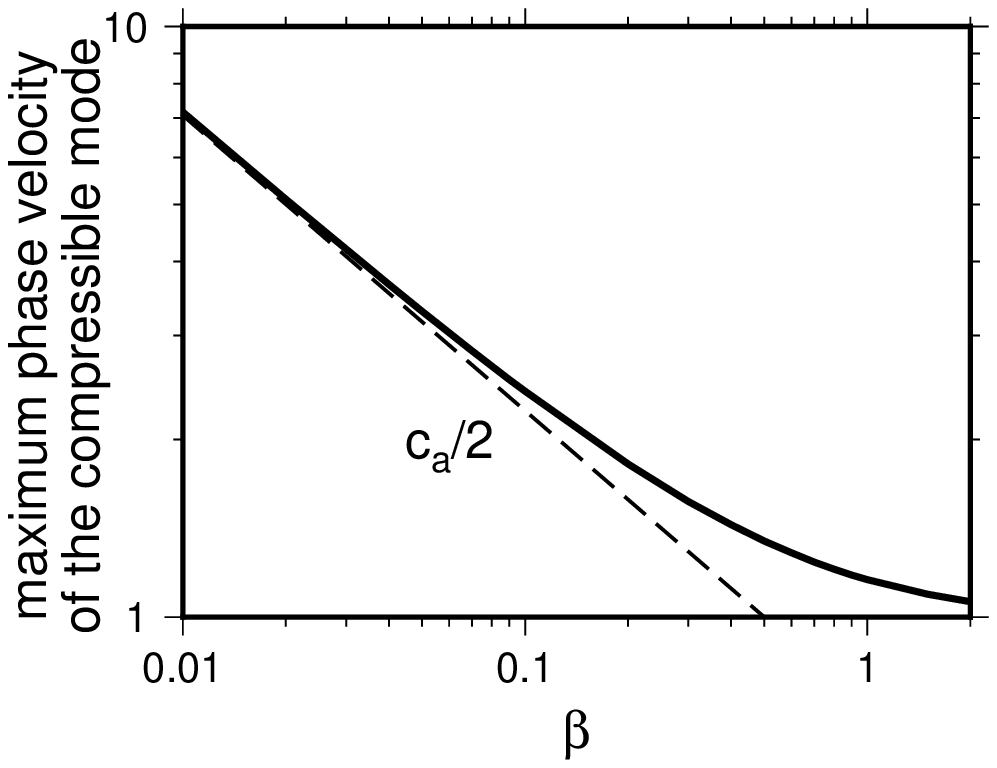}
        \end{center}
        \caption{
        The $\beta$-dependence of the maximum phase velocity of the 
        compressible mode for the 
        $\xi=1$ and $\theta=0$ case. The Gaussian kernel is used.
        The dashed line indicates $c_\mathrm{a}/2$.
        }
        \label{fig:peak}
\end{figure*}

Fig. \ref{fig:peak} shows the $\beta$-dependence of the maximum phase velocity 
of the compressible mode for $\xi=1$ and $\theta=0$. As shown in Fig. \ref{fig:cross}a, 
the ``real'' slow mode has an off-center peak. From Fig. \ref{fig:peak}, we can see that the 
maximum phase velocity increases with decreasing $\beta$.
The maximum phase velocity for low $\beta$ is well fitted by $c_\mathrm{a}/2$ shown by the dashed line 
Fig. \ref{fig:peak}. 
This dependence clearly indicates that the enhancement of 
the phase speed of the compressible wave comes from the magnetic fields.

Fig. \ref{fig:cross}b shows the results for $\xi=1/2$.
The phase velocity of the slow mode monotonically decreases with $k$.
Also for $\xi=1/2$, the mode exchange occurs at larger $k$.
From Figs.\ref{fig:cross}a and \ref{fig:cross}b, 
the extended feature around $\theta\sim 0$ in $c_\mathrm{f,num}$ in Fig. \ref{fig:kxyf}a corresponds 
to a compressible mode with a large phase velocity ($>c_\mathrm{s}$) created by mode exchange.

\section{Numerical Experiments}\label{sec:numerical}
Fig. \ref{fig:cross} reveals that the dispersion relation is abnormal if $\xi=1$ is used.
In Section \ref{sec:disp}, it is found that the compressible waves at short-wavelength 
propagate with supersonic velocities. 
In this section, 
to test this dispersive property, three simple test calculations are demonstrated.
\subsection{Propagation of an Isolated Wave}
\begin{figure*}
        \begin{center}
                \includegraphics[width=14cm]{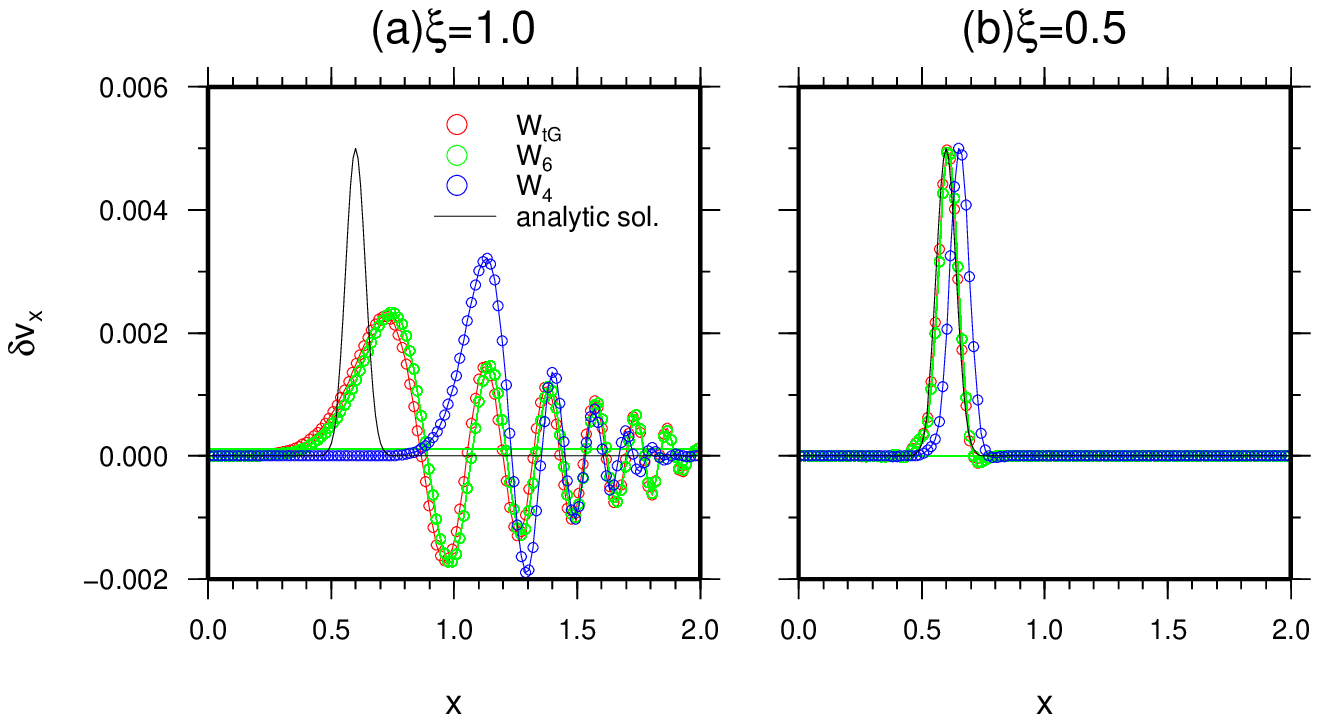}
        \end{center}
        \caption{
     Dispersion relations for (a)$\xi=1$ and (b)$\xi=1/2$.
     In each panel, the blue and red lines correspond to the fast and slow modes, respectively.
     The solid and dashed lines indicate the results with $\theta=0$ and $\theta=0.05$, 
     respectively.
        }
        \label{fig:sol}
\end{figure*}

Fist,  test of the propagation of an isolated wave is performed.
This is a severe test for the propagation of linear waves
because an isolated wave is composed of many linear waves with different wavelengths.
Thus, if the sound speed numerically has a $k$-dependence,
the shape of an isolated wave is expected to change during its propagation.
In other words, the deformation of an isolated wave clearly shows the dispersion errors. 

In the unperturbed state, the particles are distributed in a rectangular lattice in the domain 
of $[-2,2]\times[-1,1]$. The density and pressure are assumed to be 1.
A uniform magnetic field is introduced in the $x$-direction, and its strength is set
such that $\beta$ is 0.1.
A velocity perturbation given by $\delta v = 0.01 \exp(-(x/3h)^2)$ is added.
The two isolated waves propagate outward from the origin.
As the sound wave does not have dispersion, 
the two isolated waves should keep their shape with 
the sound speed ($c=1$) as they propagate.
\citet{CW03} did the same test calculation without magnetic fields.

Fig. \ref{fig:sol}a shows a snapshot of the velocity perturbation at $t=0.6$ for $\xi=1$.
Only the isolated wave propagating rightward is plotted.
The gray lines indicate the exact solution.
The colors indicate the difference of the kernel functions.
For $\xi=1$, with all kernel functions, 
the isolated waves break into smaller waves that propagates at supersonic 
velocities.
This dispersive property is consistent with the results of linear analysis.
The results with $W_6$ are almost identical to that with $W_\mathrm{tG}$. 
The isolated wave with $W_4$ shows a larger speed, as expected in Fig. \ref{fig:kxys}.
On the other hand, for $\xi=1/2$, 
there is no destruction of the isolated wave, and 
the results 
agree with the exact solution quite well, although the result with 
$W_4$ shows a slightly larger phase velocity.
This is explained by the linear analysis (see the lower left panel of Fig. \ref{fig:diag xi0.5}).
These findings suggest that the optimal choice for $\xi$ is $1/2$. Otherwise SPM provides completely 
incorrect results for the wave propagation.

\subsection{Colliding Flow Test}\label{sec:shock}
The linear analysis is valid only for the linear waves. 
In this test, we consider a colliding flow test that involves shock waves.
We investigate whether the dispersive errors affect the shock structures or not.
The computational domain is $-1<x<1$, $-0.0625<y<0.0625$, and the periodic boundary condition 
is imposed in the $y$-direction.
Initially, the density is uniform, and the gas moves toward $x=0$ from both directions with a velocity 
$\pm 3.75$. The corresponding Mach number is $4$. 
The initial magnetic field is ${\bf B}=\left(\sqrt{8\pi/\beta}, 0\right)$.
As the magnetic field is parallel to the $x$-direction, it should not affect the gas dynamics.
The truncated Gaussian kernel is used.
We consider two cases: $\beta=0.1$ and $0.01$. 
In the exact solution, two shock waves propagates outward from $x=0$.
In this test, we investigate how the strong parallel magnetic field numerically affects the gas motion.
The initial particle distribution is set to be a random distribution that is 
relaxed until the density dispersion is sufficiently small. 
In this test, a Riemann solver is used to capture shock waves.
The hyperbolic divergence cleaning method \citep{II13,II15} is also used.

\begin{figure*}
        \begin{center}
                \includegraphics[width=14cm]{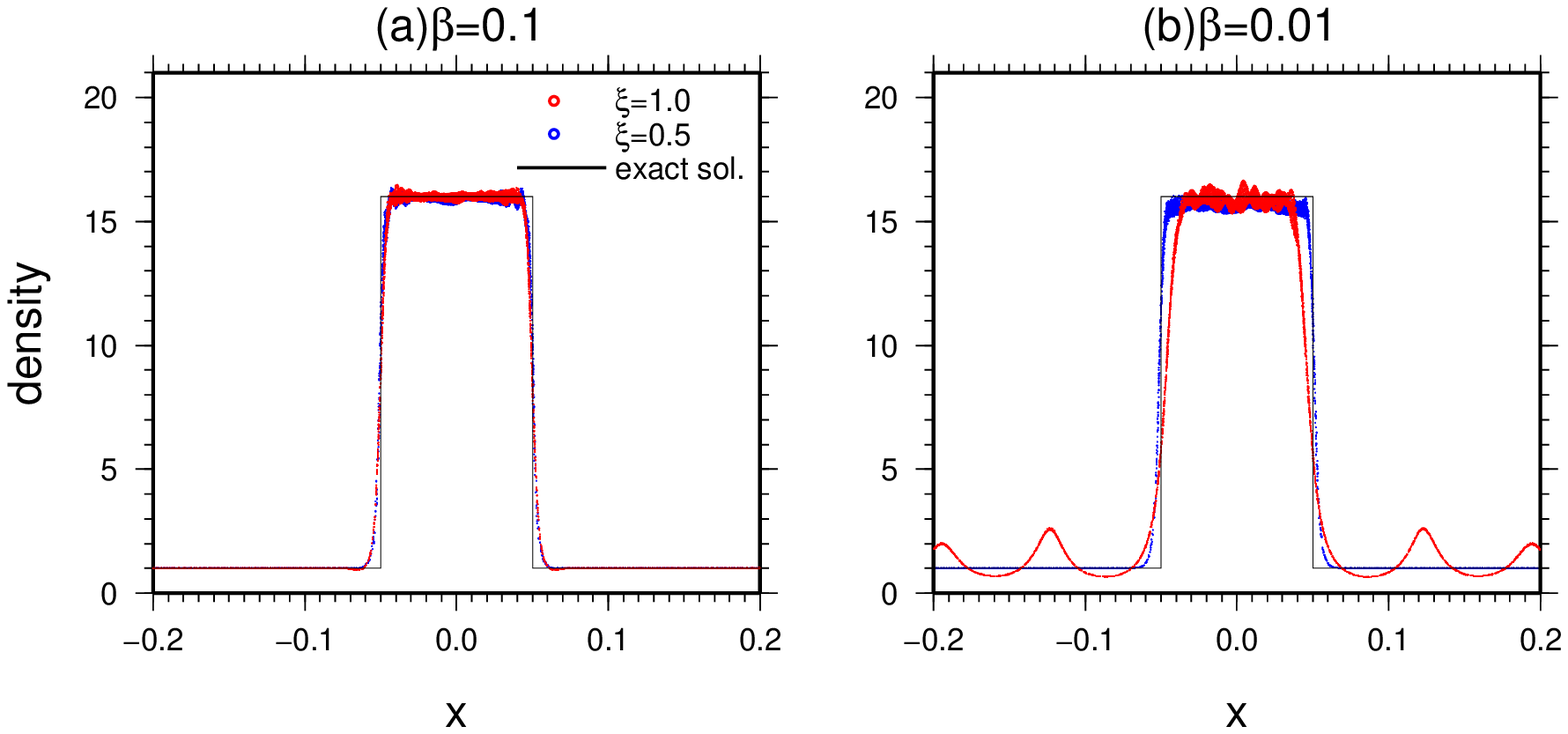}
        \end{center}
        \caption{
        Density profiles at $t=0.2$ for (a)$\beta=0.1$ and (b)$\beta=0.01$.
        Each point corresponds to an SPH particle.
     In each panel, the red and blue lines corresponds to the results with $\xi=1$ and $\xi=1/2$, respectively.
        The green lines indicate the exact solution.
        }
        \label{fig:shock}
\end{figure*}

Fig. \ref{fig:shock}a shows the results with $\xi=1$ and $\xi=1/2$ for $\beta=0.1$.
The black line indicates the exact solution. Both cases agree with the exact solution 
quite well. This is because the shock jump condition is determined only by mass and momentum conservation laws, 
regardless of their dispersive properties.
For the stronger magnetic field case $(\beta=0.01)$, the effect of the dispersive errors is shown
in Fig. \ref{fig:shock}b.
For $\xi=1/2$, the density of the shocked gas can reproduce the exact solution 
reasonably well within an error of $2\%$. 
This small error occurs due to the perpendicular magnetic field $B_y$ 
numerically generated at the shock front. 
It works as an additional pressure, leading to the smaller density jump.
For $\xi=1$, the density profile is quite different from the exact solution. 
The shock fronts broaden and small waves propagate toward the upstream with 
a supersonic velocity larger than the converging velocity
$3.75$. This can be understand by Fig. \ref{fig:peak}. 
For $\beta=0.01$, 
the maximum phase velocity of the compressible wave is $\sim c_\mathrm{a}/2 \sim 7$, which is 
larger than the converging velocity. 
Thus, waves can propagate toward 
upstream against the converging flow.

From this converging flow test, it is found that given $\beta$, 
there is a minimum shock speed below which  waves with short wavelengths 
propagate upstream and disturb the preshock gas. 
The minimum shock speed corresponds to the maximum phase velocity of the 
compressible wave, or $c_\mathrm{a}/2$ derived from the linear analysis (see Fig. \ref{fig:peak}).
In other words, if the Alfv{\'e}n Mach number is smaller than 0.5, the dispersive errors are serious.

\subsection{Hydrostatic Equilibrium Under An External Gravity}

The dispersion errors originate from the fact that a parallel magnetic field 
numerically works as an additional 
repulsive force. This indicates that the errors can be important not only for waves but also 
for hydrostatic structures.
Thus, in this test, we investigate whether SPM reproduces a hydrostatic structure under an external 
gravity, $g_x = - 2\tanh(x)$.
As the initial condition, we consider a uniform static gas with a uniform magnetic field in the $x$-direction.
The amplitude of the magnetic field is $\sqrt{8\pi/\beta}$.
The calculation domain is $-2<x<2$ and $-1/4<y<1/4$. 
The truncated Gaussian kernel is used.
A periodic boundary condition is imposed in the $y$-direction,
and the wall boundary condition is imposed in the $x$-direction.
The initial particle configuration is set to be a settled random distribution.
Because of the external gravity, the plasma accumulates toward the midplane ($x=0$).
Finally, the density profile is expected to relax to the hydrostatic 
equilibrium profile, $\rho_\mathrm{eq}(x,y)=1/\cosh(x)^2$.
To avoid an undesirable oscillation in the relaxation phase, 
we add a small drag force, $-0.005{ \bf v}/\Delta t$, 
to the equation of motion. 
If the maximum velocity becomes smaller than $0.001$, the calculations are terminated.

\begin{figure*}
        \begin{center}
                \includegraphics[width=14cm]{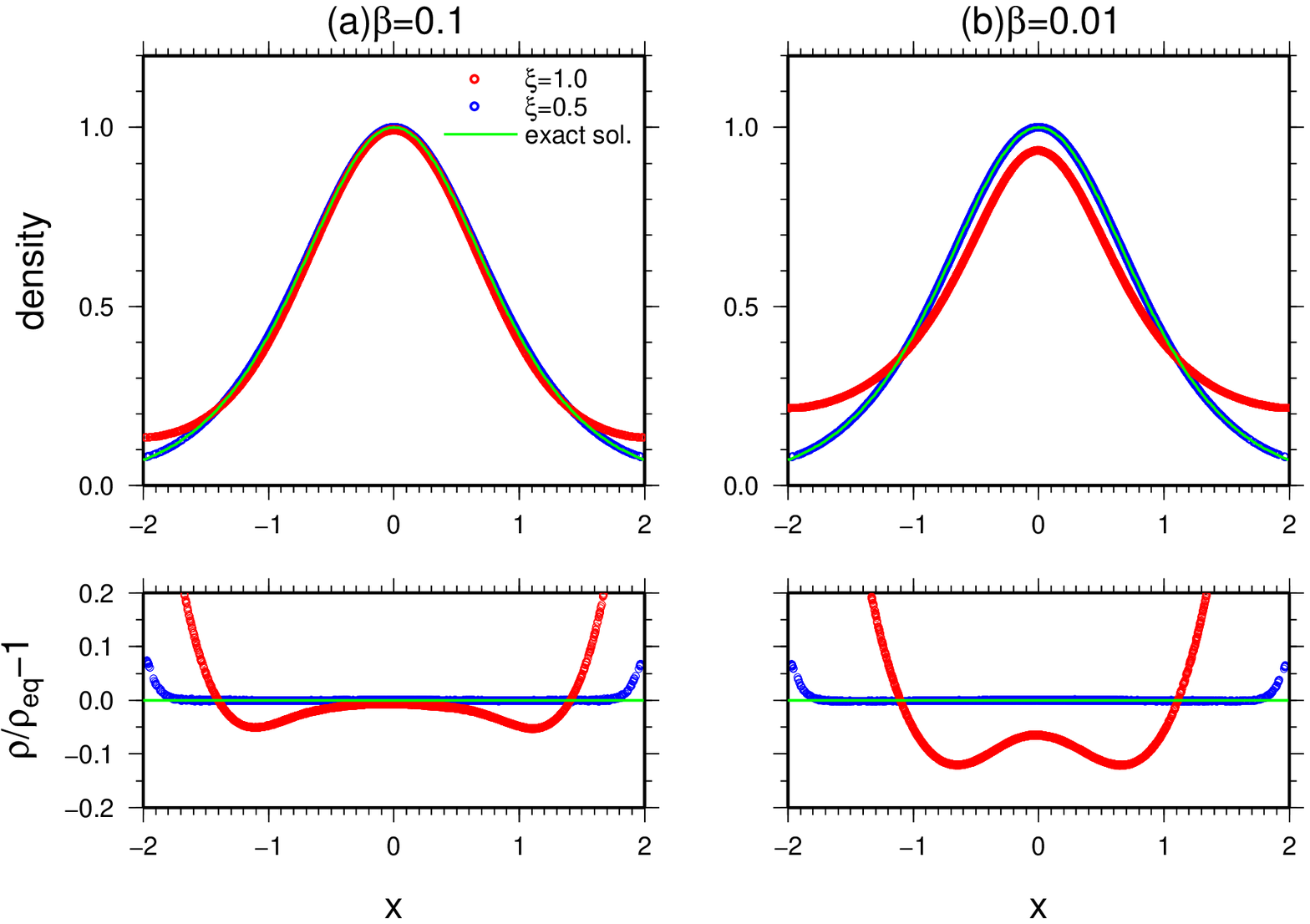}
        \end{center}
        \caption{
        {\it (Upper panels)} Density profiles for (a)$\beta=0.1$ and (b)$\beta=0.01$.
        {\it (Lower panels)} Fractional residual, $\rho(x)/\rho_\mathrm{eq}-1$.
        Each point corresponds to an SPH particle.
        In each panel, the red and blue lines corresponds to the results with $\xi=1$ and $\xi=1/2$, respectively.
        The green lines indicate the hydrostatic profile.
        }
        \label{fig:grav}
\end{figure*}

The upper panel of Fig. \ref{fig:grav}a shows the obtained density profiles for 
$\beta=0.1$. 
The lower panel of Fig. \ref{fig:grav}a indicates the fractional residual, $\rho(x)/\rho_\mathrm{eq}(x)-1$.
The result with $\xi=1/2$ agrees with the hydrostatic profile within sufficiently small error.
The small deviation around $x=\pm 2$ comes from the boundary condition.
Also for $\xi=1$, SPM reproduces the hydrostatic profile reasonably well, although 
the density in the central region is underestimated and the low density tails are overestimated.
However, this tendency becomes prominent for $\beta=0.01$, as shown in Fig. \ref{fig:grav}b.
The density profile with $\xi=1$ exhibits a more extended profile than $\rho_\mathrm{eq}(y)$.
On the other hand, even for $\beta=0.01$, SPM with $\xi=1/2$ can produce the correct profile. 
From the test, it is found that the hydrostatic profiles along 
the magnetic fields are broadened by the artificial repulsive force if $\xi=1$ is used.

\section{Discussion}\label{sec:discuss}
\subsection{Stability Against Particle Disorder}\label{sec:stable}
The test calculations demonstrated in Section \ref{sec:numerical} 
show that SPM with $\xi=1/2$ removes dispersion errors. 
However, in a realistic situation where 
a blast wave propagates in a strongly magnetized medium,
\citet{TP12} (hereafter TP12) found that SPM with $\xi<1$ produces numerical fluctuations 
behind the slow shocks and at the contact discontinuity.
They found that a value of $\xi=1$ leads to stable results.
This is because the repulsive force is weaker for
SPM with smaller $\xi$ , leading to the particle disorder.

\begin{figure*}
        \begin{center}
                \includegraphics[width=14cm]{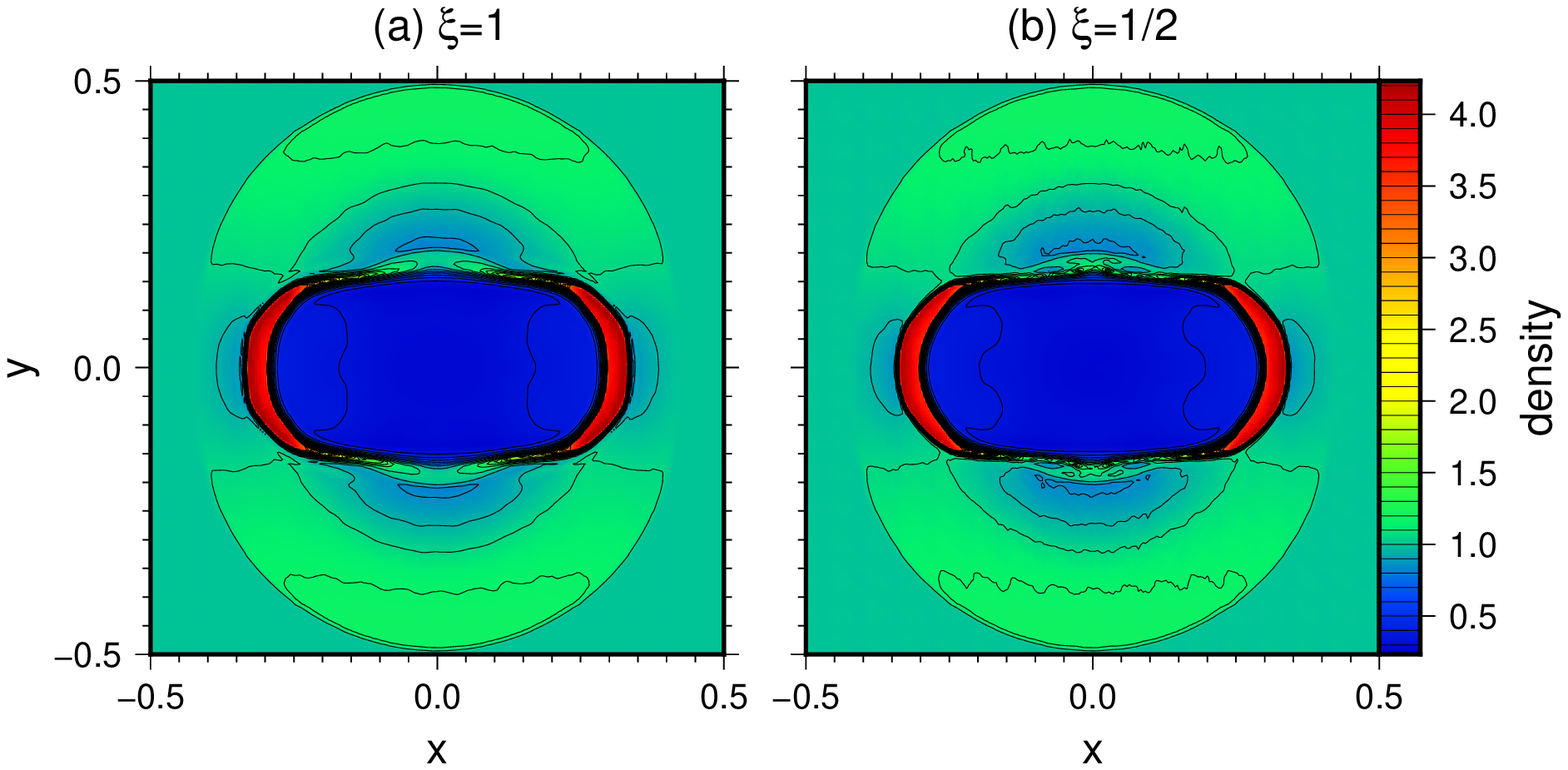}
        \end{center}
        \caption{
        Density maps of the blast wave test at $t=0.03$ for (a)$\xi=1$ 
        and (b)$\xi=1/2$.
        The truncated Gaussian kernel is used.
        A value of $\xi=1/2$ is used.
        }
        \label{fig:blast}
\end{figure*}

To investigate the stability against the particle disorder in SPM with $\xi=1/2$,
we perform the same blast wave test as in TP12.
The total particle number is $512\times 512$.
Fig. \ref{fig:blast}a and \ref{fig:blast}b show the density maps 
at $t=0.03$ for $\xi=1$ and $\xi=1/2$, respectively.
The truncated Gaussian kernel is used.
One can see that the result with $\xi=1/2$ is quite similar to that with $\xi=1$. 
As shown in Section \ref{sec:shock}, SPM can treat shock 
waves even for $\xi=1$ as long as the Alfv{\'e}n Mach number is larger than 0.5. 
Thus, it is difficult to identify the dispersion errors in this test.

Note that the significant particle disorder found by TP12
does not appear in Fig. \ref{fig:blast}b. 
Although this discrepancy may come from the difference of treatment 
of the numerical dissipations between GSPM and their scheme, 
the exact reason is still uncertain.
However, also in the GSPM,
the result with $\xi=1$ is smoother than that with $\xi=1/2$.
Thus, it is possible that GSPM with $\xi=1/2$ suffers from the serious 
particle disorder in more extreme situations.
The best choice of $\xi$ depends on situations from the point of view of
accuracy and stability.
As shown in the linear analysis,
the dispersive errors are serious in slow waves propagating along 
magnetic fields. Thus, for example, to simulate a sub-Alfv{\'e}nic turbulence in 
low $\beta$ plasma, a value of $\xi=1/2$ should be adopted.
On the other hand, in dynamical situations where strong shock waves 
are important such as the blast wave test, the dispersive errors are not serious
if the Alfv{\'e}n Mach number is larger than 0.5 (see Section \ref{sec:shock}).
In these cases, a value of $\xi=1$ is acceptable.

\subsection{Comparison with Other SPM Formulation}
\begin{figure*}
        \begin{center}
                \includegraphics[width=8cm]{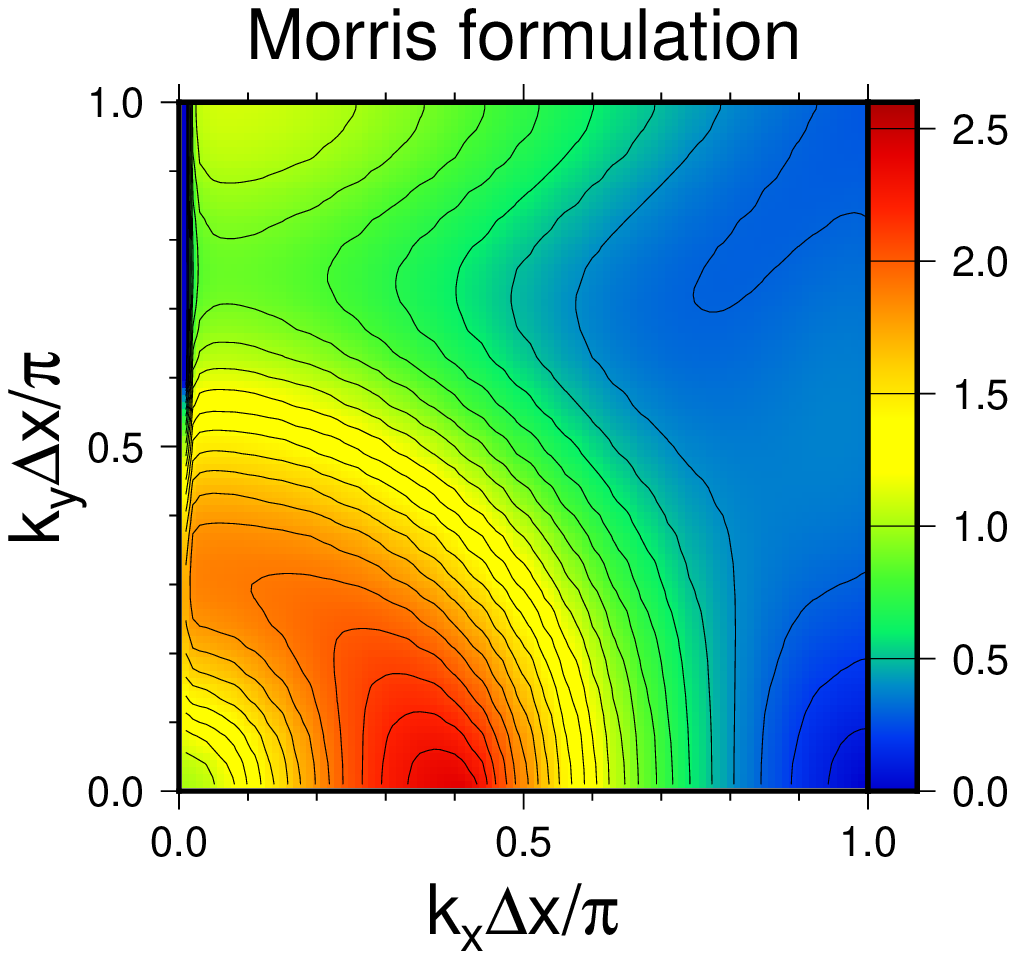}
        \end{center}
        \caption{
        Color maps of the phase velocities of the slow mode using SPM$_\mathrm{Morris}$
        in the 
        $(k_x \Delta x/\pi,k_y \Delta x/\pi)$ plane.
        The Gaussian kernel is considered.
        The parameters $\beta=0.1$ and $h=1.2\Delta x$ are used.
        The color shows the numerical phase velocity normalized by the exact phase velocity 
        depending on $\theta$.
        }
        \label{fig:kxym}
\end{figure*}

Besides the approach by \citet{Betal01}, 
\citet{M96phd} proposed the following formulation;
\begin{eqnarray}
    \frac{dv_a^\mu}{dt} &=& - \sum_b m_b 
    \left( \frac{ P_a + \frac{1}{8\pi}B_a^2}{\Omega_a\rho_a^2} 
    \nabla_a^\nu W_{ab}(h_a) + \frac{ P_b + \frac{1}{8\pi}B_b^2}{\Omega_b\rho_b^2}  
        \nabla_a^\nu W_{ab}(h_b)\right) \nonumber \\
        &+& \frac{1}{4\pi}\sum_b m_b 
        \left( \frac{B_b^\mu B_b^\nu - B_a^\mu B_a^\nu }{\rho_a\rho_b} \right)
        \frac{\nabla_a^\nu W_{ab}(h_a) + \nabla_a^\nu  W_{ab}(h_b)}{2}.
        \label{morris spm}
\end{eqnarray}
In his formulation, the conservative form is used in the isotropic part of 
the stress tensor while the non-conservative form is used in the remaining part.
Thanks to the non-conservative term, 
his formulation is free from the numerical instability.
In this section, we compare the dispersion relations between 
the Morris formulation (SPM$_\mathrm{Morris}$) and 
the conservative form with the correction term (SPM$_\mathrm{corr}$).
Linearizing equation (\ref{morris spm}), one obtains the following dispersion relation;
\begin{equation}
        \mathrm{det}\left(\omega^2 \delta^{\mu\nu} + A^{\mu\nu}\right)= 0,
        \label{disp morris}
\end{equation}
where 
\begin{eqnarray}
        A^{\mu\nu} &=& \frac{2}{\rho_0} \left(P_0 + \frac{B_0^2}{8\pi}\right) 
        \left( \phi^\mu \psi^\nu 
        -\Psi^{\mu\nu}\right) 
        - \frac{\phi^\mu}{\rho_0} \left[ P_0 \psi^\nu 
        + \frac{1}{4\pi} \left( B_0^2 \psi^\nu  
          - B_0^\eta \psi^\eta  B_0^\nu \right)\right] \nonumber \\
        &+& \frac{ B_0^\eta {\psi}^\eta}{ 4\pi}
        \left(  B_0^\zeta \psi^\zeta \delta^{\mu\nu} - B_0^\mu \psi^\nu \right).
        \label{Aij morris}
\end{eqnarray}
Fig. \ref{fig:kxym} shows the color map of the numerical phase velocities 
of the slow mode using SPM$_\mathrm{Morris}$ in the 
$(k_x\Delta x/\pi,k_y\Delta x/\pi)$ plane.
The smoothing length is assumed to be $h=1.2\Delta x$, and the plasma $\beta$ 
value is fixed to be $\beta=0.1$. The Gaussian kernel is considered.
One can see that the behavior of $c_\mathrm{s,num}$ is quite similar to that in
SPM$_\mathrm{corr}$ with $\xi=1$ (see Fig. \ref{fig:kxys}a).
The dispersion relations have an off-center peak of $c_\mathrm{s,num}$ around
$\theta \sim 0$.
Thus, the SPM formulation by \citet{M96phd} also suffers from the dispersive errors.
This behavior can be qualitatively 
understood from equation (\ref{Aij morris}).
The first term on the right-hand side of equation (\ref{Aij morris}) that 
leads to most of the dispersive errors becomes large compared with other terms for 
low $\beta$ plasma. Note that  in SPM$_\mathrm{corr}$ 
the corresponding term proportional to
$\left(\phi^\mu \psi^\nu - \Psi^{\mu\nu}\right)$
is negligible if $\xi=1/2$ is used. 

\section{Summary}\label{sec:summary}

In this study, we have investigated the dispersive properties of SPM with a correction 
term introduced to remove numerical instability in a strongly magnetized medium \citep{Betal01}.
The size of the correction term is parametrized by $\xi$ 
(see equation (\ref{eom})).
The findings in this study are summarized as follows:

\begin{itemize}
     \item As numerically found by \citet{BOT04}, the minimum value of $\xi$ 
           for removing the numerical instability is
           analytically derived as a function of the plasma $\beta$ value in Section \ref{sec:shortlimit}.

     \item For the fast modes, it is found that SPM can reproduce correct 
           phase velocities regardless of $\xi$.
           The dispersion properties are similar to those without magnetic fields.

     \item The phase velocities of the slow modes is shown to significantly depend on $\xi$. 
           For $\xi=1$, which is used in most schemes, 
           it is found that SPM suffers from significant dispersion errors with all kernel functions.
           The dispersion errors become worse for a lower value of $\beta$.
           In the long-wavelength limit $(\lambda/2\pi = 10^3\Delta x/\pi)$, 
           the numerical phase velocities are largely different from the exact 
           values especially for the cubic spline kernel. A larger smoothing length and a
           smoother kernel functions are not ultimate solution if $\beta$ is sufficiently 
           small (see Fig. \ref{fig:diag xi1.0}).
           The dispersion relations have an off-center peak of the phase velocities
           around $|k|\Delta x/\pi\sim 0.4$ and $\theta \sim 0$, where $\theta$ is 
           the angle between ${ \bf k}$ and ${ \bf B}_0$.
           Furthermore, the phase velocities are supersonic for all wavenumbers above 
           the Nyquist wavenumber $\pi/\Delta x$ (see Fig. \ref{fig:kxys}).
           The reason for this anomalous behavior is that a parallel magnetic field numerically 
           works as an additional repulsive force and 
           the phase velocities of the slow wave become supersonic.
           This fact can be understood by examining 
           the dispersion relation with respect to the slow wave propagating along a magnetic field. 
           
           On the other hand, for $\xi=1/2$, the dispersion errors found for $\xi=1$ completely disappear. 
           This can be understood analytically from the dispersion relation.

     \item To confirm the findings of the linear analysis, 
           several simple numerical experiments are demonstrated.
           For the tests of linear isolated waves and hydrostatic equilibrium, 
           SPM with $\xi=1$ leads to unphysical results while the exact solutions 
           can be reproduced for $\xi=1/2$. 
           On the other hand, SPM with $\xi=1$ can treat the parallel shock waves if Alfv{\'e}n Mach number 
   is larger than 0.5.
           This is because the shock condition is 
           determined only by 
           the conservation properties and slow waves cannot propagate toward upstream. 
           If the shock speeds are smaller, waves propagate upstream against the flow and 
           cause disturbances.  
           These results are consistent with the linear analysis.

\end{itemize}

This study clearly shows that corrected SPM with $\xi=1$ over-stabilizes the numerical instability and 
significantly modifies the dispersion properties of the slow modes.
To eliminate such dispersion errors, we suggests that $\xi=1/2$ is the best choice.
These abnormal dispersion properties have not been found in the previous works, (e.g.,
the blast wave tests used to test the capability of schemes for low $\beta$)
because, as shown in Section \ref{sec:shock}, SPM can treat shock waves even for $\xi=1$ as long 
as the shock speed is large. Thus, it is difficult to identify the dispersion errors found in this study
from the blast wave test.
The errors can be serious; for instance, in sub-Alfv{\'e}nic 
turbulence for low $\beta$ plasma, the phase velocity of the waves propagating along magnetic fields 
may be significantly overestimated.

The linear analysis and the numerical experiments suggest that 
the best choice of $\xi$ is 1/2 from the point of view of accuracy.
However, as discussed in Section \ref{sec:stable}, SPM with $\xi=1/2$
tends to less stable against particle disorder 
because of the small repulsive force.
In dynamical environments where strong shock waves 
are important, a value of $\xi=1$ is acceptable
if the Alfv{\'e}n Mach number is larger than 0.5.

\section*{Acknowledgement}
We thank the anomalous referee for valuable comments that improves this paper significantly.
We also thank Professor Shu-ichiro Inutsuka and Dr. Yusuke Tsukamoto for valuable discussions.
Numerical computations were carried out on Cray XC30 at the CfCA of National Astronomical 
Observatory of Japan.
KI was supported by the Research Fellowship from the Japan Society for 
the Promotion of Science for Young Scientist.
KI is supported by Individual Research Allowances in Doshisha University.





\end{document}